\documentclass[a4paper]{jpconf}
\usepackage{graphicx}
\usepackage{amsmath}
\usepackage{gensymb}
\usepackage{lineno}

\begin{document}
\title{Status of the SoLid experiment: Search for sterile neutrinos at the SCK$\cdot$CEN BR2 reactor}

\author{Luis Manzanillas on behalf of the SoLid collaboration}

\address{LAL, Univ Paris-Sud, CNRS/IN2P3, Universit\'e Paris-Saclay, Orsay, France}

\ead{manzanillas@lal.in2p3.fr}

\setlength{\abovecaptionskip}{5pt plus 3pt minus 2pt}
\begin{abstract}
The reactor antineutrino energy spectra and flux were reevaluated during the preparation of the recent
experiments devoted to the measurement of $\theta_{13}$. Consequently some 
discrepancies between data and the theoretical predictions in reactor antineutrino experiments at short distances were observed when using the new predicted 
flux and spectra. This problem has been called the Reactor Antineutrino Anomaly (RAA), which together with the gallium anomaly, both show discrepancies
with respect to the expectations at the $\sim$ 3 $\sigma$ level. Oscillations into a light sterile neutrino state ($\Delta m^{2} \sim 1eV^{2}$) could
account for such deficits. The SoLid experiment has been conceived to give an unambiguous response to the hypothesis of a light sterile neutrino as the 
origin of the RAA. To this end, SoLid is searching for an oscillation pattern at short baselines (6-9 m) in the energy spectrum of the $\overline{\nu}_{e}$'s
emitted by the SCK\raisebox{-0.9ex}{\scalebox{2.8}{$\cdot$}}CEN BR2 reactor in Belgium. 
The detector uses a novel technology,  combining PVT (cubes of 5$\times$5$\times$5 cm$^3$) and $^6$LiF:ZnS (sheets $\sim$ 250 $\mu$m thickness) scintillators. It is highly 
segmented (modules of 10 planes of 16$\times$16 cubes), and it's read out by a network of wavelength shifting fibers and SiPMs.
The fine segmentation and the hybrid technology of the detector allows the clear identification of 
the neutrino signals, reducing significantly backgrounds. Thus, a
high experimental sensitivity can be achieved. A 288 kg  prototype  was deployed in 2015, showing
the feasibility of the detection principle. A full scale 
detector (1.6 tons) is currently under construction, the data taking with the first detector modules 
is expected by the end of 2017. In this proceeding, the status
of the construction and the first 
results of the calibration of the first SoLid planes are presented.
\end{abstract}
\section{Introduction}
During the preparation of the recent experiments aiming the measurement of the $\theta_{13}$ mixing angle (Double Chooz, RENO and Daya Bay), 
the antineutrino flux
and energy spectra produced in nuclear reactors were re-evaluated \cite{Mueller:2011nm,Huber:2011wv}. The re-analysis of previous 
measurements in short baseline reactor antineutrino  experiments
revealed a deficit of about 6\% when comparing the data with the new predictions, and this problem has been called the 
Reactor Antineutrino Anomaly (RAA) \cite{Mention:2011rk}. Oscillations 
into a light sterile neutrino state ($\Delta m^{2} \sim 1eV^{2}$) could account for such deficit \cite{Abazajian:2012ys}. In addition, others anomalies 
pointing to the light sterile neutrino hypothesis have been observed in the past, the Gallium and the LSND anomalies \cite{Giunti:2006bj}. The RAA has
been confirmed with more recent and precise measurements of Double Chooz, Daya Bay and RENO experiments, which in addition observed a distortion in 
the anti-neutrino spectra around 5 MeV \cite{Huber:2016xis}. First hints suggest that 
this distortion is correlated to the reactor power \cite{An:2015nua} and associated to the $^{235}$U fuel \cite{An:2017osx}. 
In order to confirm or reject the light sterile neutrino hypothesis 
as the origin of the RAA, and at the same time clarify the origin of the 5 MeV distortion, new reactor neutrino experiments at short baselines are required.

In the 3+1 $\nu$ framework, the oscillation probability can be reduced to the 2 flavors approximation for a sterile neutrino at the eV scale \cite{Bilenky:1996rw}.
Then the survival probability can be written as:
\begin{equation}\label{ProbaOsc}
 P^{3+1}_{\overline{\nu}_{e}\rightarrow\overline{\nu}_{e}} 
 \simeq 1-\sin^{2}2\theta\sin^{2} \left(1.27\frac{\Delta m^{2}[eV^{2}]L[m]}{E[MeV]}\right)
\end{equation} 
where $\Delta m^{2}$ and $\theta$  are respectively the difference of squared masses $m^{2}_{4}-m^{2}_{1}$ and 
the mixing angle, both being parameters given by Nature. $E$ is 
the $\nu$ energy and $L$ is the distance traveled by the $\nu$  from the source to the detection point.
The most recent best fit values of $\theta$ and $\Delta m^{2}$ from the reactor and gallium 
anomalies \cite{Gariazzo:2017fdh} ($\Delta m^{2}\simeq 1.7$  $eV^{2}$, $\sin^{2}2\theta\simeq0.12$) implies an oscillation length
of about 4 m for $\overline{\nu}_{e}$'s of 3 MeV, which is a typical energy of detected reactor $\overline{\nu}_{e}$'s.
Thus, search for such oscillations requires a compact $\nu_{e}$ source, which can only be achieved in
research nuclear reactors or with extremely intense radioactive sources.
Furthermore, given the uncertainties in the $\overline{\nu}_{e}$ energy spectrum predictions, 
an unambiguous signature must observe the distortion of the energy spectrum as function of the distance.

A broad experimental program is ongoing to test the light sterile $\nu$ hypothesis using different detection techniques
and  $\nu$ sources. Among the experiments using a nuclear reactor as $\nu$ source we have 
the SoLid experiment \cite{Abreu:2017bpe}, whose goal is to confirm or reject the existence
of a light sterile $\nu$ state, by searching for an oscillation pattern at short baselines in the
reactor $\overline{\nu}_{e}$ energy spectrum.
The source of $\overline{\nu}_{e}$'s for the SoLid experiment is
the very compact (50 cm effective diameter) research nuclear reactor of the SCK\raisebox{-0.9ex}{\scalebox{2.8}{$\cdot$}}CEN BR2 nuclear plant, located in 
Mol, Belgium. The baselines accessible by the SoLid detector in this site varies from 6 to 9 m. This reactor uses 
highly enriched $^{235}$U, which reduces the contribution from other fissile isotopes, decreasing systematic uncertainties.  
\section{The SoLid experiment}
\begin{figure}[h]
\begin{tabular}{cc}
\includegraphics[scale=0.5]{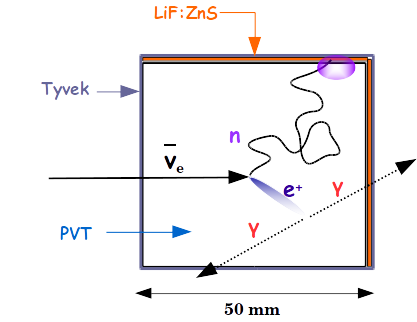}
&
\includegraphics[scale=0.3]{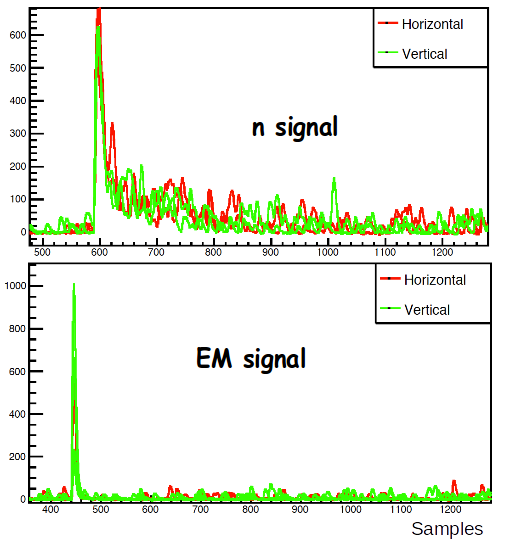}
\end{tabular}
\caption{Left: A SoLid 5$\times$5$\times$5 cm$^{3}$ cube composed of PVT and two $^{6}$LiF:ZnS(Ag) layers in two sides of the cube and wrapped with tyvek. 
Right: A characteristic neutron signal (top), which is composed of several peaks and an electromagnetic signal (bottom)}
\label{fig2}
\end{figure}

$\overline{\nu}_{e}$'s in the SoLid experiment are detected via the inverse beta decay (IBD) process: $\overline{\nu}_{e} + p \rightarrow e^{+} +n$.
The imprint of this reaction consist of a correlated prompt and delayed signals produced by the  $e^{+}$ and the $n$ respectively. These signals are
detected using a highly segmented composite detector formed of 5$\times$5$\times$5 cm$^{3}$ PVT cubes with 2 layers ($\sim$ 250 $\mu$m thickness) per cube of $^6$LiF:ZnS(Ag) 
scintillators (see figure \ref{fig2}). Each cube is wrapped with Tyvek in order to guarantee the optical isolation among cubes. PVT
is used as target for the IBD process and at the same time to detect and reconstruct the $e^{+}$ energy (prompt) which 
provides the information needed to access the $\overline{\nu}_{e}$ energy. On the other hand, 
$^6$LiF:ZnS(Ag) sheets are used to detect the neutron signals (delayed) some 
microseconds later. The neutron capture on the $^6$Li of ZnS layers produces the reaction: 
$n + ^{6}\text{Li} \rightarrow ^{3}\text{H}+\alpha +4.78 \text{ MeV}$, whose energy 
excites the phosphor grains, which slowly decays producing a very characteristic neutron signal composed of several pulses as shown in figure \ref{fig2}.

These particularities contribute to generate very different signals between neutron and electromagnetic interactions as can be observed 
in figure \ref{fig2}, which allows to achieve a good pulse shape discrimination (PSD) discrimination. 
Moreover, the neutron signals are used to set up  a dedicated neutron trigger, which can be combined with a large buffer in order to 
detect the prompt signals with a low energy threshold. 
In this way, a high IBD detection efficiency can be achieved. This neutron trigger reduces significantly backgrounds and data rate.
 The signals are read out by arrays of wavelength 
shifting fibers coupled in one side to a SiPM and a mirror in the opposite side.

The SoLid technology has been validated using several prototypes in recent years. During the 2014-2015 period, a real
scale system of 288 kg (SM1) was developed and deployed at the BR2 site. This prototype allowed to test the scalability and 
production techniques, as well as to prove the power of the segmentation. SM1 prototype showed that muon tracks can be reconstructed 
in detail, then, simple topological
cuts can be applied to reduce muon induced backgrounds. Topological cuts can be also applied to other backgrounds since the 
distance between prompt and delayed signals is only a few cubes (usually less than 3 cubes) for IBD events while it can
be larger for backgrounds. A more detailed description of the 
SoLid experiment can be found in reference \cite{Abreu:2017bpe}. 
\section{Construction and quality assurance status}
The construction of a full scale detector started in 2016 with the preparation of the frames where  arrays of 16$\times$16 cubes are 
assembled into planes, which later are joined into modules. Each SoLid module contains 10 planes. SoLid phase 1 will include 5 modules, for 
a total mass of 1.6 tons. In order to improve the neutron trigger, the SiPMs dark count rate will be reduced. To this end, 
the detector will be introduced in a container to be cooled 
at 5 $\degree$C. In addition, to reduce the reactor induced background, the container is surrounded by water walls
of 0.5 m thick and 3.4 m high (28 tons), the shielding also includes a polyethylene ceiling of 0.5 m thick (6 tons).
Finally, cadmium sheets will be placed around the container with the purpose of avoiding that neutrons coming from the outside reach the detector.
The container was already delivered in the spring of 2017, and the cooling system has been installed.
\begin{figure}[h]
\begin{tabular}{cc}
\includegraphics[scale=0.2]{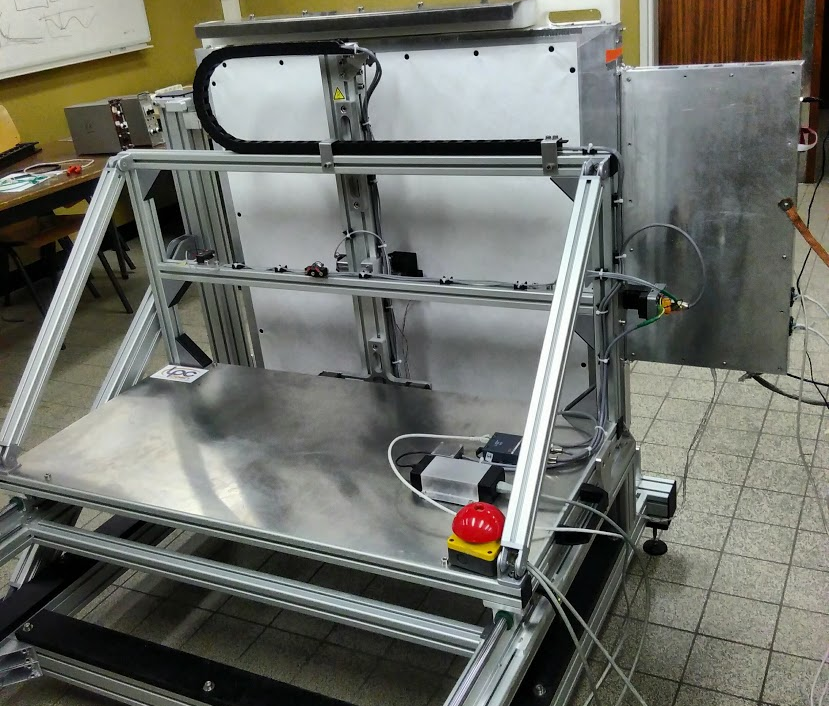}
&
\includegraphics[scale=0.3]{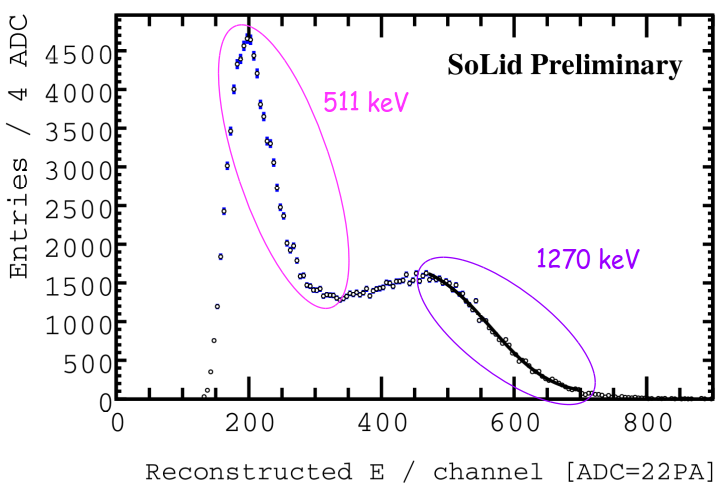}
\end{tabular}
\caption{Left: CALIPSO calibration system in $\gamma$ mode.  
Right: $^{22}$Na spectrum measured in a SoLid cube using the external trigger system.}
\label{figcalip}
\end{figure}

The totality of the PVT cubes and $^{6}$Li sheets were delivered in spring of 2017. The frame 
assembly is ongoing and will be completed by the summer of 2017. At the same time the quality assurance 
is taking place, which consist in the early detection and correction of defective 
materials or problems caused during the cube and plane productions.
To this end, an automated system has been developed which is called CALIPSO. This system can operate in
$n$, $\gamma$ and $e^{-}$ mode, and it allows to place automatically radioactive sources in front of each cube. 
In this way the light yield and the neutron capture 
efficiency can be measured in each cube individually, providing a preliminary but very good 
knowledge of the detector. Hence it has allowed a first calibration 
of the SoLid planes, and the early identification of problems
as bad couplings between fiber and SiPM, Li layers missing, dead channels, etc.

\begin{figure}[h]
\begin{tabular}{cc}
\includegraphics[scale=0.25]{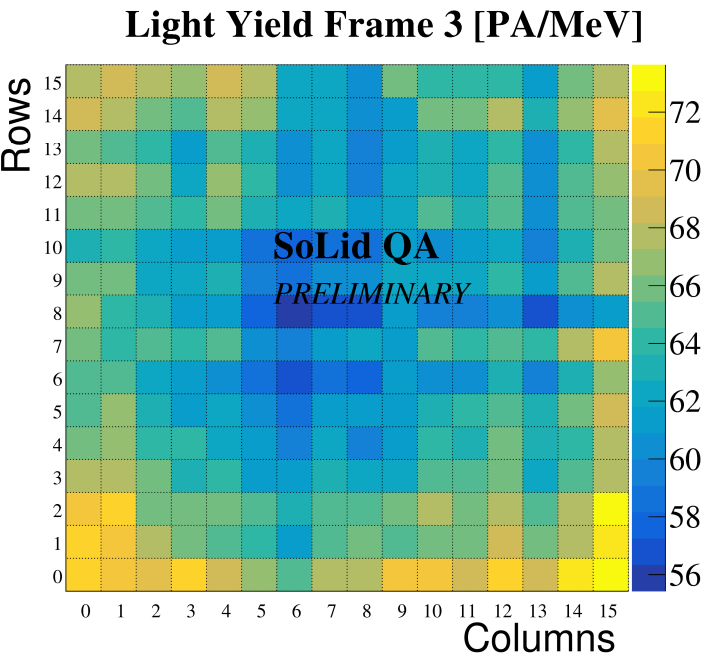}
&
\includegraphics[scale=0.17]{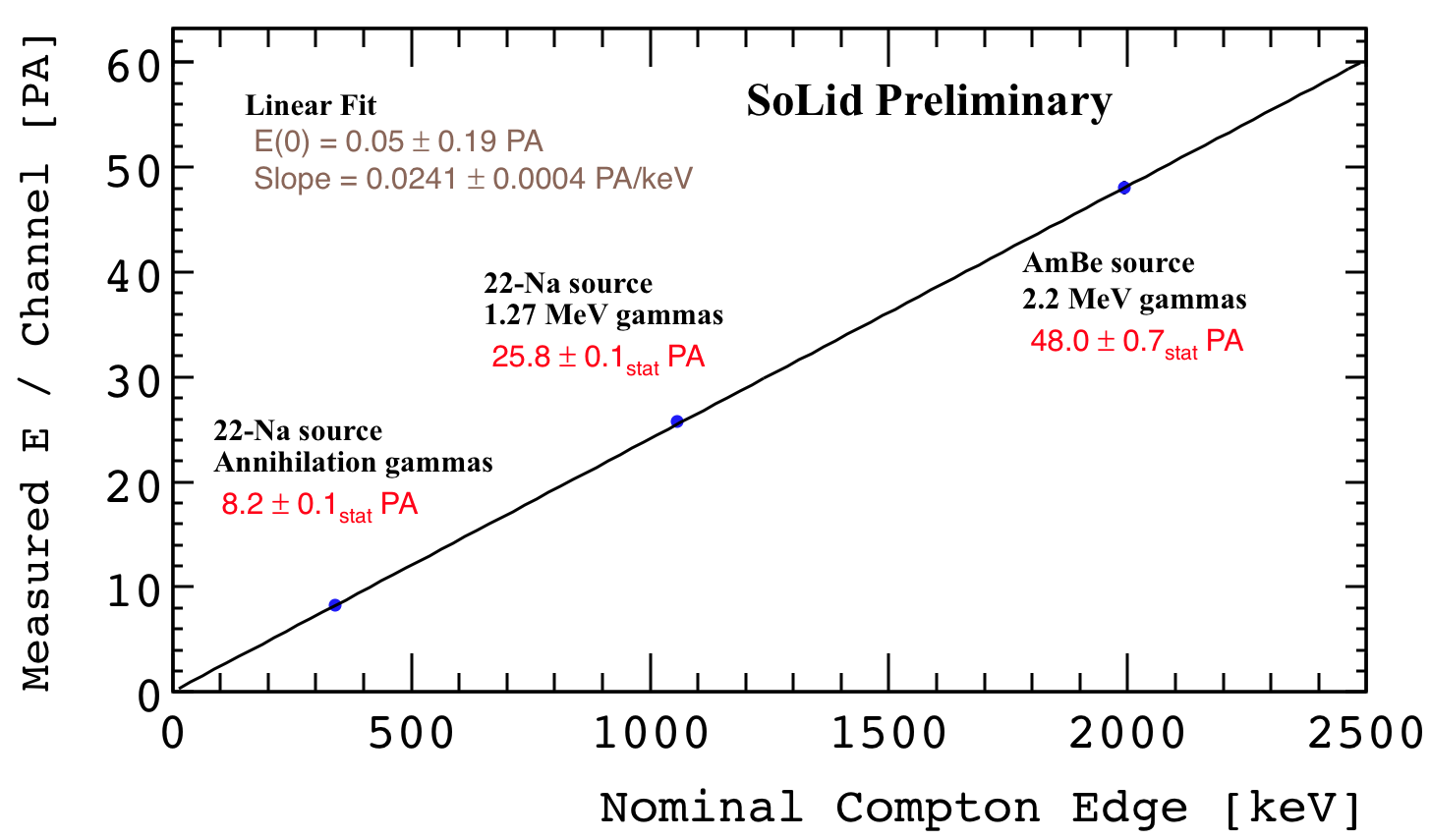}
\end{tabular}
\caption{Left: Measured light yield for each cube using a $^{22}$Na source. 
Right: PVT energy linearity response.}
\label{LYSoLid}
\end{figure}

Preliminary results confirm that SoLid planes fulfill the initial requirements of light yield and neutron capture efficiency. The 
light yield is evaluated using a $^{22}$Na gamma source, which yields back-to-back 511 keV annihilation gammas and also a 1270 keV gamma. The trigger 
is performed using an external SoLid PVT cube (5$\times$5$\times$5 cm$^{3}$) read out by a small fiber coupled to 2 SiPMs. This external trigger allows to read out the 
whole plane with almost no background, yielding ``pure'' samples for accessing the light yield, which is estimated using the compton edge 
spectrum produced by the 1270 keV gamma (see figure \ref{figcalip}). Based in a light yield optimization study 
with respect to SM1 prototype performed at LAL \cite{Boursette:2017ocg}, SoLid phase 1 was designed to have 
a light yield of at least 40 PA/MeV (16\% $\sigma_{E}/\sqrt{E}$ at 1 MeV). The first measurements
revealed that the light yield will be larger than 
60 PA/MeV (12\% $\sigma_{E}/\sqrt{E}$ at 1 MeV) (see figure \ref{LYSoLid}), exceeding the initial requirements. Using the first calibration data, the PVT energy 
linearity response was studied. Thus, 511 keV, 1270 keV gammas from $^{22}$Na and the 2220 keV from induced neutron captures on H using and AmBe neutron source were used.
Figure \ref{LYSoLid} shows a linear energy response.

On the other hand, the neutron trigger was deployed during summer of 2017.
A $^{252}$Cf and an AmBe sources were used to tune it. It is based in a peak counting algorithm, which consist in counting the number of peaks above 
a threshold in a defined time window. If a neutron capture is produced in the $^{6}$Li sheets, it will produce a waveform signal 
composed of several peaks in a few $\mu$s as showed in figure \ref{fig2}. 
The first measurement using CALIPSO system revealed a preliminary neutron capture efficiency larger than 55 \%.

\begin{figure}[h]
\begin{center}
\includegraphics[scale=0.4,trim={0 0.3cm 0 2.0cm},clip]{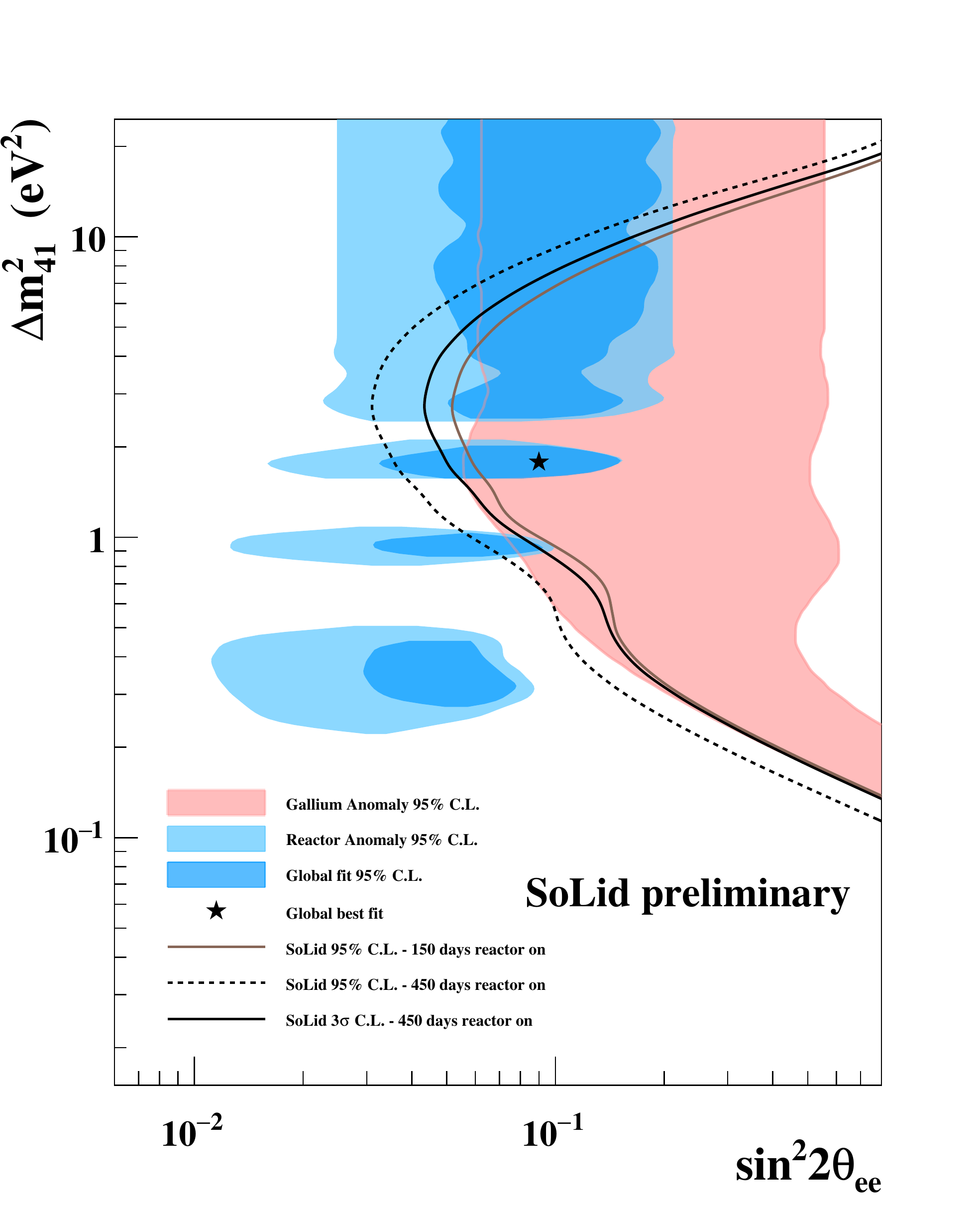}\end{center}
\caption{Expected exclusion contour of SoLid with RAA and gallium anomalies}
\label{ContourSoLid}
\end{figure}

The expected exclusion contour of the SoLid experiment is shown in figure \ref{ContourSoLid}. It covers the most interesting 
region around the best fits values of the RAA and gallium anomalies. The main assumptions of this result 
include an energy resolution of 14 \% at 1 MeV, an IBD efficiency of 30\%, a signal to background ratio of 3:1, baselines 
between 6 and 9 meters, 5 SoLid modules, and a thermal reactor power of 60 MW. Furthermore, SoLid will be able to produce 
a new electron antineutrino spectrum of reference for the $^{235}$U reactor fuel, using a new technology.
\section{Conclusions}
The SoLid collaboration has developed a new detector concept \cite{Abreu:2017bpe} based in a robust $n/e$-$\gamma$
discrimination, and with a fine segmentation for background rejection. The construction of SoLid phase 1 (1.6 tons) is ongoing,
which includes several upgrades in order to increase performances and reduce backgrounds. 
The quality assurance is taking place since the beginning of summer 2017.
The first calibration data shows that the 
light yield meets and even exceeds ($>$ 60 PA/MeV, 12\% $\sigma_{E}/\sqrt{E}$ at 1 MeV) the initial 
SoLid requirements (40 PA/MeV per cube, 16\% $\sigma_{E}/\sqrt{E}$ at 1 MeV).
On the other hand, the neutron trigger efficiency is estimated to be larger than 55\%.
Data taking is expected by autumn  2017 and the first physics results by 2018

\end{document}